\documentclass[reprint,amsmath,amssymb]{revtex4-1}

\pdfoutput=1

\usepackage[utf8]{inputenc}
\usepackage{braket,hyperref,pgfplots,tensor,tikz,float,caption,siunitx}
\usetikzlibrary{matrix}
\usepgfplotslibrary{groupplots}

\DeclareMathOperator{\Tr}{Tr}
\DeclareMathOperator{\re}{Re}
\DeclareMathOperator{\im}{Im}
\DeclareMathOperator{\sign}{sign}

\begin{document}

\title{Renormalization group approach to scalar quantum electrodynamics on de Sitter}

\author{Francisco Fabián González}
 \email{franciscofabian@me.com}
\author{Tomislav Prokopec$^\Diamond$}
 \email{T.Prokopec@uu.nl}
\affiliation{
$\Diamond$ Institute for Theoretical Physics, Spinoza Institute and EMME$\mathit{\Phi}$, Utrecht University,\\
 P.O. Box 80.089, 3508 TB, Utrecht, the Netherlands
}

\date{\today}

\begin{abstract}
We consider the quantum loop effects in scalar electrodynamics on de Sitter space 
by making use of the functional renormalization group approach.
We first integrate out the photon field, which can be done exactly to leading (zeroth) order in the gradients of the scalar field, 
thereby making this method suitable for investigating the dynamics of the infrared sector of the theory. Assuming that the scalar remains light
we then apply the functional renormalization group methods to the resulting effective scalar theory and  focus on investigating 
the effective potential, which is the leading order contribution in the gradient expansion of the effective action. 
We find symmetry restoration at a critical renormalization scale $\kappa=\kappa_{\rm cr}$ much below the Hubble scale $H$. 
When compared with the results of Serreau and Guilleux~\cite{Serreau:2014a,Guilleux:2015a} we find that the photon facilitates 
symmetry restoration such that it occurs at an RG scale $\kappa_{\rm cr}$ that is higher than in the case of a pure scalar theory. 
The true effective potential is recovered when $\kappa\rightarrow 0$ and in that limit one obtains
the results that agree with those of stochastic inflation, provided one interprets it in the sense as
advocated by Lazzari and Prokopec~\cite{Lazzari:2013a}. 
\end{abstract}

\maketitle

\section{\label{sec:1}Introduction}

   It is well known that perturbation theory for non-conformally coupled fields on de Sitter can fail.
This is because the strong interactions of these fields with gravity generate a large number of infrared modes
which may be reflected in growing secular effects in field correlators
and local observables such as mass.  For example, 
the coincident two-point function of a massless scalar exhibits on de Sitter 
a secular growth~\cite{Vilenkin:1982wt,Linde:1982uu,Starobinsky:1982ee} which for a self-interacting scalar field generates 
at two-loops a pressure and energy density that grow in time~\cite{Onemli:2002hr,Onemli:2004mb}
and a growing mass of scalars~\cite{Vilenkin:1982wt}, 
photons~\cite{Prokopec:2002jn,Prokopec:2003iu,Prokopec:2003tm} or fermions~\cite{Garbrecht:2006jm}.
Quite generically such secular effects invalidate perturbative expansion,
even if the value of the coupling constant is very small. 
In order to correctly capture observables plagued by secular effects, one ought to resort to resummation techniques. 

The best known resummation technique developed for de Sitter is Starobinsky's stochastic 
inflation~\cite{Starobinsky:1986fx}. 
This technique allows for resummation of the leading infrared effects on de Sitter 
for interacting scalar theory~\cite{Tsamis:2005hd}, Yukawa~\cite{Miao:2006pn} and quantum scalar 
electrodynamics (SQED)~\cite{Kahya:2005a,Prokopec:2007ak}. 
Remarkably, the two-loop perturbative results for the stress energy tensor~\cite{Prokopec:2006ue,Prokopec:2008a} 
agree beautifully with 
the stochastic results in the perturbative regime~\cite{Prokopec:2007ak}. But  
unlike perturbative methods, stochastic, theory allows for calculation of 
masses and energy in the late-time asymptotic regime, 
in which de Sitter symmetry gets restored~\cite{Prokopec:2007ak}. 
Furthermore, it is also known how to 
generalize stochastic inflation for interacting scalars to space-times that adiabatically deviate from de Sitter and 
to space-times of constant principal slow roll parameter $\epsilon$~\cite{Cho:2015pwa,Prokopec:2015owa}.
However, it is not known how to generalize
Starobinsky's stochastic theory to include quantum gravitational effects. Nevertheless, 
notable attempts have been made in this 
direction~\cite{Vennin:2015hra,Finelli:2010sh,Finelli:2008zg}.

It is interesting to note that one-loop perturbative calculations of the effective potential on locally de Sitter 
are available (together with quantum corrections to tensor and scalar spectral slopes and amplitudes),
as well as the one-loop effective potential induced by graviton and scalar perturbations
and a more general one-loop effective potential 
from graviton and scalar perturbations~\cite{Janssen:2008dw}.
In Ref.~\cite{Janssen:2009pb} a perturbative one-loop study of symmetry restoration in a scalar 
self-interacting theory was presented on constant $\epsilon$ space-times on which the scalar field evolves with 
the Hubble parameter; however the authors do not address what happens at late times when perturbative analysis breaks 
down.

The principal goal of works that employ resummation techniques is to overcome that limitation.
Recently, renormalization group (RG) methods have been proposed~\cite{Lazzari:2013a,Serreau:2014a,Guilleux:2015a} to
tackle these issues. 
While it is known that these methods in part resum perturbation
theory, it has not been rigorously proved that these methods indeed correctly recover the (nonperturbative) infrared physics 
in quantum field theories (QFTs) on de Sitter. 
In this work we show that functional renormalization group approach to the infrared (IR) on de Sitter, in which effective 
action is truncated at the effective potential level, leads to the results that agree with Lazzari and Prokopec 
for a self-interacting scalar field theory. More importantly, we also present here results for SQED
(in this case we do not have a stochastic inflation calculation to compare with). 

 This technique has been used from statistical physics to 
quantum gravity~\cite{Berges:2002a,Gies:2012a,Reuter:1998a,Reuter:2012id} 
and more recently it has been applied to de Sitter space dynamics~\cite{Kaya:2013a,Serreau:2014a,Guilleux:2015a}. 
The stochastic formalism has been applied with the same purpose by Lazzari and Prokopec~\cite{Lazzari:2013a}. 
In that work the authors show that infrared modes restore the symmetry in eternal inflation at late times, 
this symmetry restoration due to infrared (IR) modes was discussed previously by 
Serreau in~\cite{Serreau:2011a}. Serreau later uses the nonperturbative renormalization group 
to the same system~\cite{Serreau:2014a,Guilleux:2015a}. 
They also show that the RG flow gets dimensionally reduced to 
a zero-dimensional euclidean field theory for small curvatures (see also~\cite{Ratra:1984yq}). 
This implies that there is a symmetry restoration due to IR modes in any spacetime dimension in de Sitter. 
Serreau also shows that the late-time equilibrium state of the stochastic formalism 
is equivalent to integrating out all the IR modes using the renormalization group approach.

In the present work we begin our analysis by introducing the renormalization group techniques 
and obtaining the RG flow equation in section \ref{sec:2}. In section \ref{sec:3} we present the SQED theory 
in de Sitter and calculate its effective potential and energy-momentum tensor. 
We show the flow of a Higgs-like potential with photons interactions coming from SQED 
on top of it in section \ref{sec:3-1}, and in section \ref{sec:3-2} we compute the flow of the SQED stress-energy tensor. 
The final conclusions and outlook are presented in section \ref{sec:4}. 
In appendix \ref{sec:a} we discuss the flow of a scalar field theory without photons interactions, 
to compare how photons affect the flow of the scalar field. There we show that photons enhance symmetry restoration.

\section{\label{sec:2}General setup for studying RG flow}

We begin with a scalar field theory on de Sitter space in $D=d+1$ spacetime dimensions whose tree-level action is,
\begin{equation}
S[\varphi]=\int d^{D}x\sqrt{-g}\left\{-\frac{1}{2}g^{\mu\nu}\partial_{\mu}\varphi(x)\partial_{\nu}\varphi(x)-V(\varphi)\right\},
\end{equation}
with the metric tensor defined by,
\begin{equation}
\mathrm{d}s^{2}=a(\eta)^{2}\big(\!-\mathrm{d}\eta^{2}+\mathrm{d}\vec{x}^{\,2}\big)
\,,
\end{equation}
where $a(\eta)=-1/(H\eta)$, using conformal time, $-\infty<\eta<0$, and comoving spatial coordinates, $\vec{x}$,
which is related to the physical distance $\vec{x}_{\rm phys}$ by, $d\vec{x}_{\rm phys}=a(\eta)d\vec{x}$. 

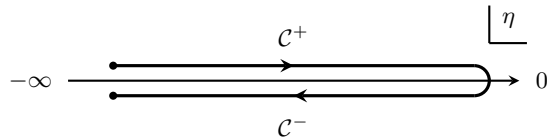
\begin{figure}[H]
\centering
\begin{tikzpicture}[line width=1.2pt,>=stealth]
\draw[->,thick] (0,1) -- (6,1);
\node[fill,circle,scale=0.35] (S) at (0.6,1.2) {};
\node[fill,circle,scale=0.35] (F) at (0.6,0.8) {};
\draw[->] (0.6,1.2) -- (3,1.2);
\draw (2.9,1.2) -- (5.4,1.2);
\draw (5.4,1.2) arc (90:-90:0.2cm);
\draw[->] (5.4,0.8) -- (3,0.8);
\draw (3.1,0.8) -- (0.6,0.8);
\draw[thick] (5.6,1.5) -- (6.1,1.5);
\draw[thick] (5.6,1.5) -- (5.6,2.0);
\draw[very thick] (5.86,1.8) node {$\eta$};
\draw[very thick] (6.3,1.0) node {$0$};
\draw[very thick] (-0.5,1.0) node {$-\infty$};
\draw[very thick] (3,1.6) node {$\mathcal{C}^{+}$};
\draw[very thick] (3,0.4) node {$\mathcal{C}^{-}$};
\end{tikzpicture}
\caption{Closed time path $\mathcal{C}=\mathcal{C}^{+}\cup\mathcal{C}^{-}$. The upper branch $\mathcal{C}^{+}$ goes forward from the initial to the final time and the lower branch $\mathcal{C}^{-}$ goes backward from the final to the initial time.}
\label{fig:path}
\end{figure}

Here we are interested in studying quantum effects in the non-equilibrium setting of an inflationary universe for which 
the Schwinger-Keldysh formalism~\cite{Schwinger:1960qe,Keldysh:1964ud}
is the method of choice. 
This formalism is suitable for calculation of (time dependent) expectation values of 
operators~\cite{Calzetta:1987a,Calzetta:1988a,Berges:2005a,Berges:2012a} as well as for calculation of the effective action. 
In that formalism it is useful to define the complex integration contour ${\cal C}$~\cite{Schwinger:1960qe}
shown in figure~\ref{fig:path}.
The corresponding non-equilibrium effective action is then defined as a Legendre transform
of $W_{\cal C}[J]=-\imath \ln(Z_{\cal C}[J])$, where $Z_{\cal C}[J]$ is the generating functional defined by,
\begin{equation}
Z_{\cal C}[J]=e^{iW_{\cal C}[J]}=\int\mathcal{D}\varphi\ e^{i\left(S_{\cal C}[\varphi]+\int_{\cal C}J(x)\varphi(x)\right)}
\,,
\label{generating functional}
\end{equation}
which is associated with the tree-level action along the contour ${\cal C}$, 
\begin{equation}
S_{\cal C}[\varphi]=\int_x 
           \left\{-\frac{1}{2}g^{\mu\nu}\partial_{\mu}\varphi(x)\partial_{\nu}\varphi(x)-V(\varphi)\right\}
\,,
\label{action along C}
\end{equation}
where 
\begin{equation}
\int_x\equiv \int_{\cal C} d^{D}x\sqrt{-g}
\label{contour integration: def}
\end{equation}
implies that one integrates in time along the contour ${\cal C}$ in figure~\ref{fig:path}.

The vacuum expectation value of the field $\varphi$ is the relevant quantity in studying the spontaneous symmetry breaking of the theory and is defined as~\cite{Jordan:1986a}
\begin{equation}
\frac{1}{\sqrt{-g(x)}}\frac{\delta W_{\cal C}[J]}{\delta J(x)}=\Braket{\varphi(x)}_{J}\equiv\phi(x),
\end{equation}
We are dealing here with a massless scalar on de Sitter, and trying to impose 
the Chernikov-Tagirov (also known as the Bunch-Davies) vacuum~\cite{Chernikov:1968zm}
would lead to infrared divergences of correlation functions~\cite{Allen:1987tz}.
The correct procedure is to regulate the infrared in a manner that deviates from 
the Chernikov-Tagirov vacuum state. The regularization can be done either by matching onto 
a pre-inflationary phase in which the vacuum is regular in the infrared~\cite{Janssen:2009nz}, 
or by placing the Universe in a comoving box~\cite{Janssen:2009pb}.
One can also show that -- when expanded in inverse powers of 
the box size $L$ -- the latter regularization is equivalent to imposing an infrared comoving momentum cutoff, $k_0\sim 1/L$.
One can show that any regularization procedure (including those mentioned above) leads to correlators 
that exhibit a breaking of de Sitter symmetry~\cite{Janssen:2009pb,Miao:2010a}. 
In this paper we regulate the infrared (IR) by introducing an infrared regulator $R_{\kappa}$ 
and define the average effective action as a modified Legendre transform,
\begin{equation}
\Gamma_{\kappa}[\phi]=W_{\kappa}[J]-\int_{x}J(x)\phi(x)-\frac{1}{2}\int_{x,y}R_{\kappa}(x,y)\phi(x)\phi(y)
\,,
\end{equation}
where $W_{\kappa}[J]$ is defined as
\begin{equation}
e^{iW_{\kappa}[J]}=\int\mathcal{D}\varphi\ e^{i\left(S_{\cal C}[\varphi]+\int_{x}J(x)\varphi(x)+\frac{1}{2}
\int_{x,y}R_{\kappa}(x,y)\varphi(x)\varphi(y)\right)}
\,.
\end{equation}
The regulator $R_{\kappa}$ acts as a momentum-dependent mass term that is finite for IR modes ($p<\kappa$) 
and vanishes for UV modes ($p\gg \kappa$)~\cite{Kaya:2013a}, such that one recovers the full effective action 
when the cutoff is removed, 
\begin{equation}
\Gamma_{\kappa=0}[\phi]=\Gamma[\phi]
\,, 
\label{effective action}
\end{equation}
and one recovers the tree-level (bare) action in the opposite limit, 
$\Gamma_{\kappa\rightarrow \infty}[\phi]=S_{\cal C}[\phi]$. We choose the Litim regulator~\cite{Litim:2001a}
\begin{equation}
R_{\kappa}(p)=Z_{\kappa}\left(\kappa^{2}-(k/a)^{2}\right)\theta\left(\kappa^{2}-(k/a)^{2}\right)
\,,\label{eq:litimreg}
\end{equation}
where $\sqrt{Z_{\kappa}}$ is a renormalization factor for the scalar field $\phi$ that can depend 
on the moment cutoff $\kappa$, $k$ is the comoving momentum and $a=a(\eta)$ the scale factor.

We use the local potential approximation (LPA)~\cite{Wetterich:1993a} as an {\it Ansatz} for the average effective action,
\begin{equation}
\Gamma_{\kappa}[\phi]=\int_{x}\left\{-\frac{Z_{\kappa}}{2}g^{\mu\nu}\partial_{\mu}\phi(x)\partial_{\nu}\phi(x)
                -V_{\kappa}(\phi)\right\}
\,,
\end{equation}
which should be good for the infrared sector of the theory on de Sitter because the large Hubble damping in de Sitter 
tends to suppress higher-derivative terms in the effective action (an analogous approximation is used in 
stochastic inflation~\cite{Lazzari:2013a}). 

The evolution with the running of the cutoff of the average effective action is given by the Wetterich equation~\cite{Wetterich:1993b}
\begin{equation}
\dot{\Gamma}_{\kappa}[\phi]=\frac{1}{2}\Tr\left\{\dot{R}_{\kappa}(x;y)G_{\kappa}(y;x)\right\},
\label{Wetterich equation}
\end{equation}
where we have defined the notation $\dot{F}\equiv\kappa\partial_{\kappa}F$, $\Tr\equiv\int_{x,y}$ 
and $G_{\kappa}(x,y)$ is the (full) propagator of the average effective action,
\begin{equation}
iG_{\kappa}(x;y)^{-1}
  =\frac{1}{\sqrt{g(x)g(y)}}\frac{\delta^{2}\Gamma_{\kappa}[\phi]}{\delta\phi(y)\delta\phi(x)}+R_{\kappa}(x;y).
\label{Green function}
\end{equation}

We begin our analysis by deriving the flow equation for the effective potential and for small $\kappa$ ($\kappa<H$), 
which is the relevant regime for infrared effects. Namely, at momentum scales $k/a>\kappa$ we assume 
that the theory is perturbative, such that the effective action at the scale $\kappa$ can be approximated by 
its local (tree-level) form; the quantum effects from the running from some high scale $\kappa_0\gg H$ 
up to $\kappa\lesssim H$ can be to a good approximation subsumed to a finite renormalization of the coupling parameters 
in $V(\phi)$ and $Z_\kappa$. 

 Next step is to transform from real to momentum space, 
which can be done by performing a Wigner transform (which is a Fourier transform with respect to the relative coordinate), 
so using isotropy and homogeneity of the spatial coordinates the regulator in momentum representation is
\begin{equation}
\begin{gathered}
R_{\kappa}(x;y)=R_{\kappa}(\eta,\eta';\|\vec{x}-\vec{y}\|)\\
=\int\frac{\mathrm{d}^{d}k}{(2\pi)^{d}}e^{i\vec{k}\cdot(\vec{x}-\vec{y})}\tilde{R}_{\kappa}\left(\eta,\eta';k\right),
\end{gathered}
\end{equation}
where $\vec{k}$ is the comoving spatial momentum, $k=\|\vec k\|$ and we have introduced the notation $\tilde{F}$
for the Wigner transform of $F$.

Requiring de Sitter symmetry implies that the Feynman propagator function can depend only on the 
invariant distance $y(x,x')$~\cite{Chernikov:1968zm,Parentani:2013a}
\begin{equation}
y(x;x')\equiv a(\eta)a(\eta')H^{2}\left(\!\|\vec{x}\!-\!\vec{x}'\|^{2}\!-\!\left(|\eta\!-\!\eta'|\!-\!i\epsilon\right)^{2}\right),
\label{invariant distance y}
\end{equation}
which is related to the geodesic distance $\ell(x;x')$ as, $y(x;x')_{\epsilon\rightarrow 0}=4\sin^2\big(H\ell(x;x')\big)$.
The $i\epsilon$-prescription in~(\ref{invariant distance y}) provides the correct boundary conditions for the 
Feynman propagator.
Next, it is convenient to perform a conformal rescaling of all the quantities,
\begin{equation}
\phi(x)\rightarrow a(\eta)^{\frac{d-1}{2}}\phi(x)=(-\eta H)^{-\frac{d-1}{2}}\phi(x),
\end{equation}
and more generally,
\begin{equation}
\begin{gathered}
F(x;x')\rightarrow\left(a(\eta)a(\eta')\right)^{d_{F}}F(\eta,\eta';\|\vec{x}\!-\!\vec{x}'\|)\\
=\left(H^{2}\eta\eta'\right)^{-d_{F}}F(\eta,\eta';\|\vec{x}\!-\!\vec{x}'\|)
\,,
\end{gathered}
\end{equation}
where $d_{F}$ is the conformal dimension of the quantity $F$. In the momentum $p$-representation $p=-kH\eta$, 
$p'=-kH\eta'$ and introducing the notation $\hat{F}$ for dimensionless functions, {\it i.e.} 
extracting all the scale and dimensional factors, the regulator becomes,
\begin{equation}
\tilde{R}_{\kappa}\left(\eta,\eta';k\right)=\left(H^{2}\eta\eta'\right)^{\frac{d+3}{2}}k^{3}\hat{R}_{\kappa}\left(p;p'\right),
\end{equation}
\begin{equation}
\hat{R}_{\kappa}(p;p')=H\frac{\delta_{\mathcal{C}}\left(p-p'\right)}{p^{2}}R_{\kappa}(p),
\end{equation}
and the Green function is
\begin{equation}
\tilde{G}_{\kappa}\left(\eta,\eta';k\right)=\frac{\left(H^{2}\eta\eta'\right)^{\frac{d-1}{2}}}{k}\hat{G}_{\kappa}\left(p;p'\right).
\end{equation}
We are primarily interested in the flow of the effective potential in the infrared. For 
that reason we can evaluate the Wetterich equation~(\ref{Wetterich equation})
at a constant field $\phi$ (neglecting the derivatives 
of $\phi$ is, namely, justified in the infrared because of the Hubble dumping),
\begin{equation}
\left.\dot{\Gamma}_{\kappa}[\phi]\right|_{\phi=\textrm{const.}}=\left.\int_{x}\left\{-\dot{V}_{\kappa}(\phi)\right\}\right|_{\phi=\textrm{const.}},
\label{Gamma: EPA}
\end{equation}
the integral over $x$ is just a volume factor $\Omega\equiv\int_{x}$.
Upon inserting Eq.~(\ref{Gamma: EPA}) into~(\ref{Wetterich equation}) after some algebra 
and making use of
$\dot{V}_{\kappa}(\phi)=\left.-\Omega^{-1}\dot{\Gamma}_{\kappa}[\phi]\right|_{\phi=\textrm{const.}}$, one obtains~\cite{Serreau:2014a},
\begin{equation}
\begin{gathered}
\dot{V}_{\kappa}(\phi)=-\frac{1}{2\Omega}\Tr\left\{\dot{R}_{\kappa}(x;y)G_{\kappa}(y;x)\right\}\\
=\frac{1}{2}\int\frac{\mathrm{d}^{d}p}{(2\pi)^{d}}\,\dot{R}_{\kappa}
            \left(p\right)\frac{\hat{F}_{\kappa}\left(p;p\right)}{p},
\end{gathered}
\end{equation}
where the Green function $G_{\kappa}(y;x)$ is defined in Eq.~(\ref{Green function}). 
By making use of the generating function the Green function can be also defined as the two-point function 
ordered along the complex time contour,
\begin{equation}
\begin{gathered}
G_{\kappa}(x;x')=\Braket{T_{\mathcal{C}}\phi(x)\phi(x')}\\
=F_{\kappa}(x;x')-\frac{i}{2}\sign_{\mathcal{C}}(\eta-\eta')\rho_{\kappa}(x;x'),
\end{gathered}
\end{equation}
where we have defined 
the Hadamard (statistical) two-point function $F_{\kappa}(x;x')\equiv\frac{1}{2}\Braket{\left\{\phi(x),\phi(x')\right\}}$ 
and the Pauli-Jordan or Schwinger (spectral) two-point function 
$\rho_{\kappa}(x;x')\equiv i\Braket{\left[\phi(x),\phi(x')\right]}$~\cite{Berges:2005a,Berges:2012a}. 
The Feynman propagator satisfies the following equation,
\begin{equation}
\begin{gathered}
\left(\partial_{p}^{2}+\frac{1}{H^{2}}-\left(\nu_{\kappa}^{2}-\frac{1}{4}
-\frac{R_{\kappa}(p)}{Z_{\kappa}H^{2}}\right)\frac{1}{p^{2}}\right)\hat{G}_{\kappa}\left(p,p'\right)\\
   =\frac{i\delta_{\mathcal{C}}\left(p-p'\right)}{Z_{\kappa}H}
,\label{eq:green1}
\end{gathered}
\end{equation}
with
\begin{equation}
\nu_{\kappa}=\sqrt{\frac{d^{2}}{4}-\frac{V''_{\kappa}(\phi)}{Z_{\kappa}H^{2}}}.
\end{equation}
Following~\cite{Serreau:2014a} we assume factorization of the two-point functions, 
\begin{equation}
 \hat G_\kappa(p;p') =  \hat F_\kappa(p;p') - \frac{i}{2}\sign_{\mathcal{C}}(\eta\!-\!\eta')\hat \rho_{\kappa}(p;p'),
\label{hat G kappa}
\end{equation}
in terms of a new function $u_{\kappa}(p)$ that only depends on one momentum $p$, 
\begin{eqnarray}
\hat{F}_{\kappa}(p;p') \!&=&\! Z_{\kappa}^{-1}\re\left\{u_{\kappa}(p)u_{\kappa}^{*}(p')\right\}
\nonumber\\
\hat{\rho}_{\kappa}(p;p') \!&=&\! -2Z_{\kappa}^{-1}\im\left\{u_{\kappa}(p)u_{\kappa}^{*}(p')\right\}
\,.
\label{factorization F kappa}
\end{eqnarray}
This is possible only if $\hat G_\kappa(p;p')$ describes a pure state, which is what we assume in this work.
Upon doing that and  inserting the expression for the Litim regulator~\eqref{eq:litimreg}, 
the equation for the Green's function~\eqref{eq:green1} is
\begin{equation}
\left[\partial_{p}^{2}+\frac{1}{H^{2}}\!-\!\left(\!\nu_{\kappa}^{2}\!-\!\frac{1}{4}\right)\frac{1}{p^{2}}\right]u_{\kappa}(p)=0,
\quad {\rm for}\;\, p\ge\kappa
\end{equation}
%
and
\begin{equation}
\left[\partial_{p}^{2}\!-\!\left(\!\bar\nu_{\kappa}^{2}\!-\!\frac{1}{4}\right)\frac{1}{p^{2}}\right]u_{\kappa}(p)=0,
\quad {\rm for}\;\, p\le\kappa
\,,
\end{equation}
%
where we have defined 
\begin{equation}
\bar{\nu}_{\kappa}=\sqrt{\nu_{\kappa}^{2}-\frac{\kappa^{2}}{H^{2}}}
\,.
\label{bar nu: def}
\end{equation} 
The properly normalized solutions 
({\it i.e.} with a canonically normalized Wronskian) of those differential equations are~\cite{Janssen:2009pb,Kaya:2013a},
\begin{equation}
u_{\kappa}(p)=\sqrt{\frac{\pi p}{4H}}e^{i\gamma_{\kappa}}H^{(1)}_{\nu_{\kappa}}
                   \left(\frac{p}{H}\right)\qquad(p\ge\kappa)
\,,
\label{Mode function UV}
\end{equation}
and
\begin{equation}
u_{\kappa}(p)=\sqrt{\frac{\pi p}{4H}}e^{i\gamma_{\kappa}}
    \left[c_{\kappa}^{-}\frac{\kappa^{\bar\nu_{\kappa}}}{p^{\bar\nu_{\kappa}}}
                             +c_{\kappa}^{+}\frac{p^{\bar\nu_{\kappa}}}{\kappa^{\bar\nu_{\kappa}}}\right]\qquad(p\le\kappa)
\,,
\end{equation}
where $H^{(1)}_{\nu_{\kappa}}(z)$ is the Hankel function of the first kind 
and $\gamma_{\kappa}=\frac{\pi}{2}\left(\nu_{\kappa}+\frac{1}{2}\right)$. 
Note that~(\ref{Mode function UV}) is a mode function that corresponds to the Chernikov-Tagirov vacuum of a 
particle whose mass-squred is $V''(\phi)>0$. 
Requiring continuity of the function and its first derivative we obtain
\begin{equation}
c_{\kappa}^{\pm}=\frac{1}{2}\left[H^{(1)}_{\nu_{\kappa}}\left(\frac{\kappa}{H}\right)\pm\frac{\kappa}{\bar\nu_{\kappa}}\frac{d}{dp}H^{(1)}_{\nu_{\kappa}}\left(\frac{\kappa}{H}\right)\right].
\end{equation}
The flow of the effective potential as a function of $u_{\kappa}(p)$ is,
\begin{equation}
\dot{V}_{\kappa}(\phi)=\frac{\Omega_{d}}{2(2\pi)^{d}}
\!\int_{0}^{\kappa}\!\mathrm{d}p\,p^{d-2}
      \left[\left(2-\eta_{\kappa}\right)\kappa^{2}+\eta_{\kappa}p^{2}\right]\left|u_{\kappa}(p)\right|^{2}
,
\label{the flow of Vkappa}
\end{equation}
with $\eta_{\kappa}=-\dot{Z}_{\kappa}/Z_{\kappa}$ and $\Omega_{d}=2\pi^{d/2}/\Gamma(d/2)$. Expanding $u_{\kappa}(p)$ for small $\kappa$ using,
\begin{equation}
H_{\nu_{\kappa}}(\kappa)=\frac{2^{\nu_{\kappa}}\Gamma(\nu_{\kappa})}{i\pi \kappa^{\nu_{\kappa}}}\left(1+\mathcal{O}\left(\kappa^{2}\right)\right)\qquad\left(\kappa^{2}\ll 1\right),
\end{equation}
the flow equation for the effective potential becomes,
\begin{equation}
\dot{V}_{\kappa}(\phi)=\frac{\Omega_{d}F_{\nu_{\kappa}}}{2(2\pi)^{d}}\frac{\kappa^{d+2-2\nu_{\kappa}}}{H^{1-2\nu_{\kappa}}}\left[\frac{2-\eta_{\kappa}}{d-2\bar\nu_{\kappa}}+\frac{\eta_{\kappa}}{d+2-2\bar\nu_{\kappa}}\right],
\end{equation}
where we have defined,
\begin{equation}
F_{\nu_{\kappa}}=\frac{\left(2^{\nu_{\kappa}}\Gamma(\nu_{\kappa})\right)^{2}}{4\pi}.
\end{equation}
We make one last approximation, $|V_{\kappa}''(\phi)|\ll H^2$, and take $Z_{\kappa}=1$ to obtain 
the final expression for the flow of the effective potential~\cite{Serreau:2014a,Guilleux:2015a},
\begin{equation}
\dot{V}_{\kappa}(\phi)=2A_{d}\frac{\kappa^{2}H^{d+1}}{V''_{\kappa}(\phi)+\kappa^{2}},
\label{flow for effective potential}
\end{equation}
with
\begin{equation}
A_{d}=\frac{d\Omega_{d}F_{d/2}}{4(2\pi)^{d}}=\frac{\Gamma(\frac{d}{2}+1)}{4\pi^{d/2+1}}
\,.
\end{equation}
The flow equation~(\ref{flow for effective potential}) is the main result of this section, and we use it in the rest of the paper
to study the flow of the effective action in the infrared on de Sitter.

\section{\label{sec:3}Flow of the effective potential}

In this section we study the flow of a spontaneously broken potential with photons interactions using the general flow equation obtained in the previous section. We first obtain the SQED effective potential integrating out the vector potential and add it on top of a Higgs potential.
Finally we also study the flow of the SQED stress-energy tensor.

\subsection{\label{sec:3-1}Scalar QED}

The starting point is the SQED action for a massless, minimally coupled (MMC) charged scalar
\begin{eqnarray}
S_{\rm SQED}\!&=&\!\int\! d^Dx\sqrt{-g}\Big\{\!-g^{\mu\nu}\left(D_{\mu}\Phi\right)^{\dag}\!\left(D_{\nu}\Phi\right)
\nonumber\\
&&\hskip 1.9cm
-\,\frac{1}{4}g^{\mu\nu}g^{\rho\sigma}\tilde F_{\mu\rho}\tilde F_{\nu\sigma}\!\Big\}
\,,
\end{eqnarray}
where $D_{\mu}=\partial_{\mu}-ie\tilde A_{\mu}$ is a covariant derivative and
$\tilde F_{\mu\nu}=\partial_{\mu}\tilde A_{\nu}-\partial_{\nu}\tilde A_{\mu}
  =\nabla_{\mu}\tilde A_{\nu}-\nabla_{\nu}\tilde A_{\mu}$
is the field strength associated with the gauge field $\tilde A_\mu$. 
Since under the gauge transformations the fields transform as,
\begin{eqnarray}
 \tilde A_\mu \!&\rightarrow&\! \tilde A_\mu - \partial_\mu \Lambda(x)
\nonumber\\
\Phi \!&\rightarrow&\! e^{-ie\Lambda(x)}\Phi 
\,,
\label{gauge transformations}
\end{eqnarray}
the following redefinition of the fields 
\begin{eqnarray}
\Phi(x) \!&=&\! \frac{\phi(x)}{\sqrt{2}}e^{i\theta(x)}
\nonumber\\
A_{\mu}(x)  \!&=&\! \tilde A_{\mu}(x)-\frac1e\partial_{\mu}\theta(x)
\label{Phi A decomposition}
\end{eqnarray}
results in an action which can be expressed solely in terms of gauge invariant quantities $\phi$ and $A_\mu$ 
which does not transform under the gauge transformations~(\ref{gauge transformations}). The resulting action is,  
\begin{equation}
\begin{gathered}
S_{\rm SQED}=\int_{x}\left\{-\frac{1}{2}g^{\mu\nu}\partial_{\mu}\phi\partial_{\nu}\phi
                  -\frac{1}{2}e^{2}g^{\mu\nu}A_{\mu}A_{\nu}\phi^{2}\right.\\
                     \left.
\hskip 1.5cm
-\frac{1}{4}g^{\mu\nu}g^{\rho\sigma}F_{\mu\rho}F_{\nu\sigma}
                            -\frac{1}{2}\delta\xi\phi^{2}R-\frac{1}{4}\delta\lambda\phi^{4}\right\}
,
\label{eq:sqedbare}
\end{gathered}
\end{equation}
where $F_{\mu\nu}=\partial_{\mu}A_{\nu}-\partial_{\nu}A_{\mu}=\tilde F_{\mu\nu}$. The last equality holds because of 
the gauge invariance of $F_{\mu\nu}$. Anticipating dimensional regularisation and renormalization procedure, 
we have added in~(\ref{eq:sqedbare}) two local counterterms (whose couplings are $\delta\xi$ and $\delta\lambda$) 
which we will need to renormalize 
the photon fluctuations on de Sitter.

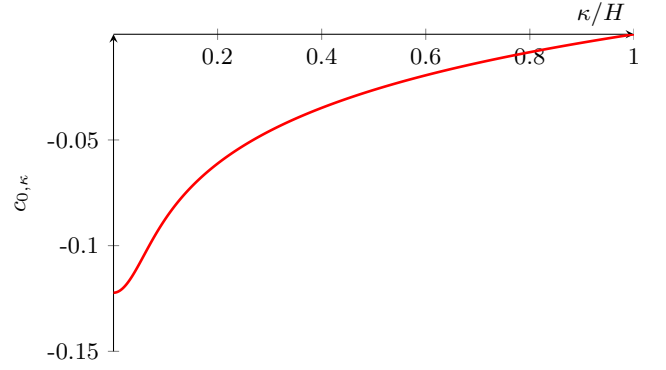
\begin{figure}[H]
\begin{center}
\begin{tikzpicture}
\begin{axis}[axis x line=center,axis y line=left,ylabel near ticks,ytick={-0.05,-0.1,-0.15},yticklabels={-0.05,-0.1,-0.15},width=8.5cm,height=5.8cm,xlabel={$\kappa/H$},ylabel={$c_{0,\kappa}$},xmin=0,xmax=1,ymin=-0.15,ymax=0.0]
 \addplot[color=red,smooth,line width=1.0pt] table {con0_0.0001.dat};
\end{axis}
\end{tikzpicture}
\end{center}
\caption{Running of  $c_{0,\kappa}$, the constant contribution to $V_\kappa$ in Eq.~(\ref{eq:effpotansatz}--\ref{eq:effpotansatz:2}).}
\label{fig:run1s_extra}
\end{figure}

\begin{figure}[H]
\begin{center}
\begin{tikzpicture}
\begin{axis}[axis x line=center,axis y line=left,ylabel near ticks,width=8.5cm,height=5.8cm,xlabel={$\kappa/H$},
ylabel={$c_{2,\kappa}$},xmin=0,xmax=1,ymin=-0.00015,ymax=0.003]
 \addplot[color=red,smooth,line width=1.0pt] table {con2_0.0001.dat};
\end{axis}
\end{tikzpicture}
\end{center}
\caption{\small Running of the quadratic coupling constant $c_{2,\kappa}$ (mass term) 
defined in~(\ref{eq:effpotansatz}--\ref{eq:effpotansatz:2}). 
$c_{2,\kappa}$ flips sign at $\kappa=\kappa_{\rm cr}\simeq 0.184H$, 
where the symmetry restoration occurs.}
\label{fig:run1s}
\end{figure}
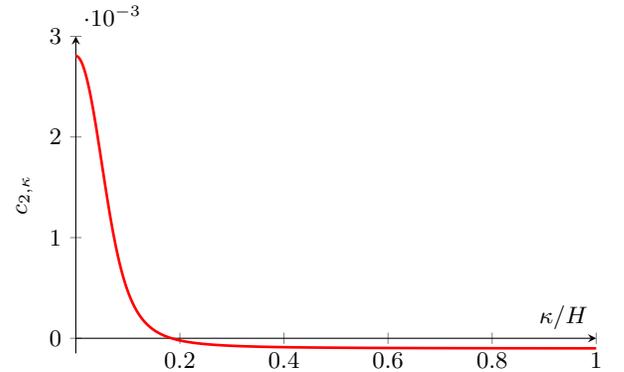

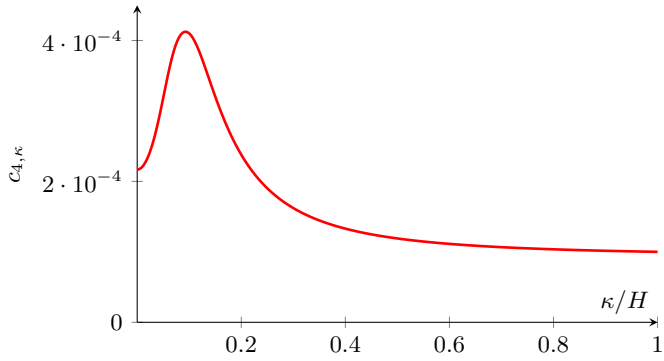
\begin{figure}[H]
\begin{center}
\begin{tikzpicture}
\begin{axis}[axis x line=center,axis y line=left,ylabel near ticks,width=8.5cm,height=5.8cm,xlabel={$\kappa/H$},
ylabel={$c_{4,\kappa}$},xmin=0,xmax=1,ymin=0,ymax=0.00045]
 \addplot[color=red,smooth,line width=1.0pt] table {con4_0.0001.dat};
\end{axis}
\end{tikzpicture}
\end{center}
\caption{Running of the quartic coupling constant $c_{4,\kappa}$ (scalar self-interaction)
 defined in~(\ref{eq:effpotansatz}--\ref{eq:effpotansatz:2}).}
\label{fig:run2s}
\end{figure}

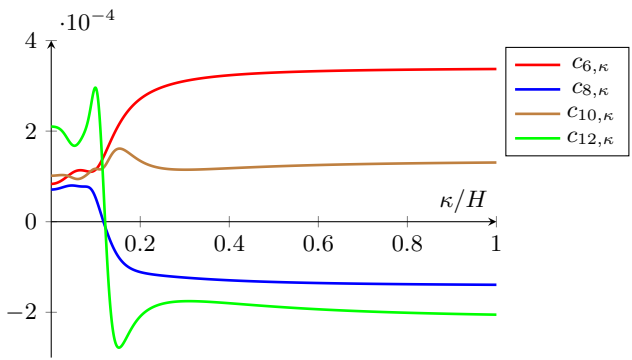
\begin{figure}[H]
\begin{center}
\begin{tikzpicture}
\begin{axis}[legend style={at={(1.02,0.98)},anchor=north west},axis x line=center,axis y line=left,ylabel near ticks,width=7.5cm,height=5.8cm,xlabel={$\kappa/H$},xmin=0,xmax=1,ymin=-0.0003,ymax=0.0004]
 \addplot[color=red,smooth,line width=1.0pt] table {con6_0.0001.dat};
 \addplot[color=blue,smooth,line width=1.0pt] table {con8_0.0001.dat};
 \addplot[color=brown,smooth,line width=1.0pt] table {con10_0.0001.dat};
 \addplot[color=green,smooth,line width=1.0pt] table {con12_0.0001.dat};
 \legend{$c_{6,\kappa}$,$c_{8,\kappa}$,$c_{10,\kappa}$,$c_{12,\kappa}$}
\end{axis}
\end{tikzpicture}
\end{center}
\caption{Running of the coupling constants $c_{6,\kappa}$, $c_{8,\kappa}$, $c_{10,\kappa}$ and $c_{12,\kappa}$
 defined in~(\ref{eq:effpotansatz}--\ref{eq:effpotansatz:2}).}
\label{fig:run3s}
\end{figure}

\begin{figure}[H]
\begin{center}
\begin{tikzpicture}
\begin{axis}[legend style={at={(0.60,0.98)},anchor=north west},axis x line=center,axis y line=left,ylabel near ticks,width=8.5cm,height=5.8cm,xlabel={$\kappa/H$},xmin=0,xmax=1,ymin=-0.004,ymax=0.009]
 \addplot[color=red,smooth,line width=1.0pt] table {con14_0.0001.dat};
 \addplot[color=blue,smooth,line width=1.0pt] table {con16_0.0001.dat};
 \addplot[color=brown,smooth,line width=1.0pt] table {con18_0.0001.dat};
 \legend{$c_{14,\kappa}$,$c_{16,\kappa}$,$0.1\times c_{18,\kappa}$}
\end{axis}
\end{tikzpicture}
\end{center}
\caption{\small Running of the coupling constants $c_{14,\kappa}$, $c_{16,\kappa}$ and $c_{18,\kappa}$
 defined in~(\ref{eq:effpotansatz}--\ref{eq:effpotansatz:2}).}
\label{fig:run4s}
\end{figure}
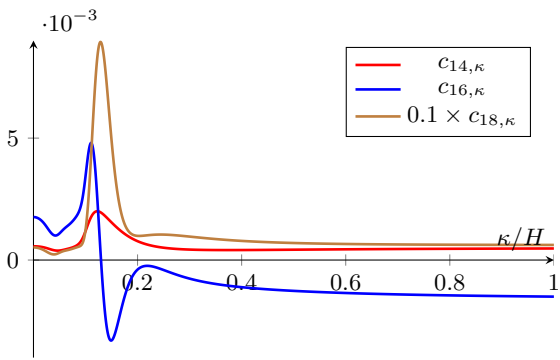

The next step is to integrate out the vector field to obtain an effective action for the 
scalar field~\cite{Prokopec:2007ak,Prokopec:2008b}, 
\begin{equation}
e^{i\Gamma[\phi]}=\int\mathcal{D}A_{\mu}  \delta(\nabla^\alpha A_\alpha) e^{iS_{\rm SQED}}
\label{integrating vector field}
\end{equation}
where the $\delta$-function imposes the exact Lorenz gauge condition, $\nabla^\alpha A_\alpha=0$,
in the functional integral measure. As we argue below, this is the right condition to impose since 
the photon field we integrate over becomes massive due to the backreaction of long wavelength scalar fluctuations on de Sitter.
Formally, upon integrating the photons, the effective action for scalars~(\ref{integrating vector field}) becomes,
\begin{equation}
\begin{gathered}
\Gamma[\phi]=\int_{x}\left\{-\frac{1}{2}g^{\mu\nu}\partial_{\mu}\phi\partial_{\nu}\phi-\frac{1}{2}\delta\xi\phi^{2}R-\frac{1}{4}\delta\lambda\phi^{4}\right\}\\
+\frac{i}{2}\log\bigg\{\det\left[\partial_{\mu}\left(\sqrt{-g}\left(g^{\mu\nu}g^{\rho\sigma}
                    \!-\!g^{\mu\sigma}g^{\rho\nu}\right)\partial_{\nu}\right)\right.\\
\left.\hskip -1.5cm
-\,e^{2}\sqrt{-g}g^{\rho\sigma}\phi^{2}\right]\bigg\},
\end{gathered}
\label{effective action: potential approximation}
\end{equation}
where the operator inside the determinant is the inverse of the photon propagator,
\begin{equation}
\begin{gathered}
iG^{\mu\nu}(x,x')^{-1}=\frac{1}{\sqrt{g(x)g(x')}}\frac{\delta^{2}S}{\delta A_{\nu}(x')\delta A_{\mu}(x)}\\
\hskip 1cm
  =\frac{1}{\sqrt{-g}}\bigg\{\frac{1}{\sqrt{-g}}\partial_{\alpha}\left(\sqrt{-g}\left(g^{\alpha\sigma}g^{\mu\nu}-g^{\alpha\mu}g^{\nu\sigma}\right)\partial_{\sigma}\right)\\
\hskip 0cm
-e^{2}\phi^{2}(x)g^{\mu\nu}\bigg\}\delta_{\mathcal{C}}(x-x')
.
\end{gathered}
\label{photon propagator}
\end{equation}
Since the scale at which the scalar field in~(\ref{photon propagator}) varies is super-Hubble which --
upon making the approximation $e^{2}\phi^{2}(x) =  m^{2}_{\gamma}+{\rm gradient\;\;corrections}$
-- has been calculated to leading order in gradients by Tsamis and Woodard in~\cite{Tsamis:2007a}, 
\begin{equation}
\begin{gathered}
iG_{\mu\nu}(x,x)=g_{\mu\nu}\frac{d}{2}\frac{H^{2}}{m^{2}_{\gamma}}\frac{H^{d-1}}{(4\pi)^{(d+1)/2}}\left(\frac{\Gamma\left(d\right)}{\Gamma\left(\frac{d+1}{2}+1\right)}\right.\\
\left.-\Gamma\left(-\frac{d+1}{2}\right)\frac{\Gamma\left(\frac{d+2}{2}+\nu\right)\Gamma\left(\frac{d+2}{2}-\nu\right)}{\Gamma\left(\frac{1}{2}+\nu\right)\Gamma\left(\frac{1}{2}-\nu\right)}\right)
,
\end{gathered}
\end{equation}
with
\begin{equation}
\nu=\sqrt{\left(\frac{d-2}{2}\right)^{\!2}-\frac{m_{\gamma}^{2}}{H^{2}}}
.
\end{equation}
By differentiating the effective action~(\ref{effective action: potential approximation})
 one obtains the derivative of the effective potential,
\begin{equation}
\begin{gathered}
\frac{\mathrm{d}V_{\textrm{eff}}\left(\phi^{2}\right)}{\mathrm{d}(\phi^2)}=\frac{1}{2}\delta\xi H^{2}d(d+1)+\frac{1}{2}\delta\lambda\phi^{2}\\
+\frac{e^{2}}{2}d(d+1)\frac{H^{2}}{2m^{2}_{\gamma}}\frac{H^{d-1}}{(4\pi)^{(d+1)/2}}\left[\frac{\Gamma\left(d\right)}{\Gamma\left(\frac{d+1}{2}+1\right)}\right.\\
\left.-\Gamma\left(-\frac{d+1}{2}\right)\frac{\Gamma\left(\frac{d+2}{2}+\nu\right)\Gamma\left(\frac{d+2}{2}-\nu\right)}{\Gamma\left(\frac{1}{2}+\nu\right)\Gamma\left(\frac{1}{2}-\nu\right)}\right].
\end{gathered}
\end{equation}
Next, upon performing dimensional regularization~\cite{Hooft:1972a} the divergent and finite part 
are~\cite{Prokopec:2008b},
\begin{equation}
\begin{gathered}
\frac{\mathrm{d}V_{\textrm{eff}}\left(\phi^{2}\right)}{\mathrm{d}(\phi^2)}
   =\frac{1}{2}\delta\xi H^{2}d(d+1)+\frac{1}{2}\delta\lambda\phi^{2}\\
+\frac{e^{2}}{4}d(d+1)\frac{H^{d-1}}{(4\pi)^{(d+1)/2}}\left\{\frac{2}{d\!-\!3}
     \left(1\!+\!\frac{e^{2}\phi^{2}}{2H^{2}}\right)\!+\!\frac{1}{2}\right.\\
\left.+\left(1\!+\!\frac{e^{2}\phi^{2}}{2H^{2}}\right)\left[\Psi\left(\frac{3}{2}\!+\!\nu\right)
             \!+\!\Psi\left(\frac{3}{2}\!-\!\nu\right)\!-\!\frac{3}{2}\!+\!\gamma\right]\right\}\\
+\mathcal{O}(d\!-\!3)
,
\end{gathered}
\end{equation}
where $\Psi(x)$ is the digamma function. We choose the following counterterms to cancel the divergencies and the $\phi^{2}$ and $\phi^{4}$ terms of the effective potential,
\begin{equation}
\delta\xi=\frac{e^{2}H^{d-3}}{(4\pi)^{(d+1)/2}}\left(\!-\frac{1}{d\!-\!3}+\frac{\gamma}{2}+\mathcal{O}(d\!-\!3)\right),\label{eq:cont1}
\end{equation}
and
\begin{equation}
\delta\lambda=\frac{d(d+1)e^{4}H^{d-3}}{2(4\pi)^{(d+1)/2}}\left(\!-\frac{1}{d\!-\!3}\!+\!\frac{\gamma}{2}
        \!-\!\frac{3}{4}\!+\!\mathcal{O}(d\!-\!3)\right).
\label{eq:cont2}
\end{equation}
It is useful to define $\displaystyle z\equiv e^{2}\phi^{2}/2H^{2}=m^{2}_{\gamma}/2H^{2}$ and integrate with respect to $z$, taking the limit $\epsilon\rightarrow 0$ the effective potential is,
\begin{equation}
\begin{gathered}
V_{\textrm{eff}}(z)=\frac{3H^{4}}{8\pi^{2}}\left\{\left(-1+2\gamma\right)z+\left(-\frac{3}{2}+\frac{\gamma}{2}\right)z^{2}\right.\\
+\int_{0}^{z}\mathrm{d}y\ (1+y)\left(\Psi\left(\frac{3}{2}+\frac{1}{2}\sqrt{1-8y}\right)\right.\\
\left.\left.+\Psi\left(\frac{3}{2}-\frac{1}{2}\sqrt{1-8y}\right)\right)\right\}.
\end{gathered}
\label{Veff of z}
\end{equation}
 There is no simple expression for $V_{\textrm{eff}}(z)$ since one cannot express the integral in~(\ref{Veff of z}) 
in terms of known functions. There is a relatively simple expansion of $V_{\textrm{eff}}(z)$ around $z=0$.
To accomplish that, note first that, 
\begin{equation}
  x(y)\equiv \frac{1}{2}-\frac{1}{2}\sqrt{1-8y}=\sum_{n=0}^{\infty}\frac{(2n)!2^{n+1}}{n!(n+1)!}y^{n+1},
\end{equation}
and then,
\begin{eqnarray}
V_{\textrm{eff}}(z)\!&=&\!\frac{3H^{4}}{8\pi^{2}}\Bigg\{-z^{2}
\label{Veff of z:2}\\
&&\hskip -1.3cm
-\,\int_{0}^{z}\mathrm{d}y\,(1\!+\!y)\left(2\sum_{n=1}^{\infty}\zeta(2n+1)x(y)^{2n}\!-\!\sum_{n=1}^{\infty}x(y)^{n}\right)
\Bigg\}.
\nonumber
\end{eqnarray}
One can easily perform the integrals in~(\ref{Veff of z:2}).  Taking account of $z=e^{2}\phi^{2}/2H^{2}$ 
and keeping terms up to order $\phi^{18}$ we get a good approximation for small values of the field, 
$|\phi|\lesssim 6.5 H^{2}$, and the effective scalar field potential for SQED is,
\begin{equation}
\begin{gathered}
V_{\textrm{eff}}\left(\phi\right)=\frac{3H^{4}}{8\pi^{2}}\left\{\left[10-8\zeta(3)\right]\left(\frac{e^{2}}{2H^{2}}\right)^{3}\frac{\phi^{6}}{3}\right.\\
\left.+\left[12-10\zeta(3)\right]\left(\frac{e^{2}}{2H^{2}}\right)^{4}\phi^{8}\right.\\
+\left[264-192\zeta(3)-32\zeta(5)\right]\left(\frac{e^{2}}{2H^{2}}\right)^{5}\frac{\phi^{10}}{5}\\
+\left[1568-1056\zeta(3)-288\zeta(5)\right]\left(\frac{e^{2}}{2H^{2}}\right)^{6}\frac{\phi^{12}}{6}\\
+\left[9792-6272\zeta(3)-2048\zeta(5)-128\zeta(7)\right]\left(\frac{e^{2}}{2H^{2}}\right)^{7}\frac{\phi^{14}}{7}\\
+\left[7920-4896\zeta(3)-1760\zeta(5)
      -208\zeta(7)\right]\left(\frac{e^{2}}{2H^{2}}\right)^{8}\phi^{16}\\
+\left[420992-253440\zeta(3)-96768\zeta(5)-15360\zeta(7)\right.\\
\left.\left.-512\zeta(9)\right]\left(\frac{e^{2}}{2H^{2}}\right)^{9}\frac{\phi^{18}}{9}
+\mathcal{O}\left(\phi^{20}\right)\right\}.\label{eq:sqedeffpot}
\end{gathered}
\end{equation}
Our next task is to add this potential on top of the Higgs potential and calculate the running 
of all the parameters. To do that we choose the following {\it Ansatz} for the effective scalar potential at a scale $\kappa$,
\begin{equation}
\begin{gathered}
V_{\kappa}(\phi)=H^4\sum_{n=0}^{\infty}\frac{c_{2n,\kappa}}{(2n)! }\left(\frac{\phi^2}{H^2}\right)^{n}
,
\label{eq:effpotansatz}
\end{gathered}
\end{equation}
where the coupling constants are defined as
\begin{equation}
c_{2n,\kappa}=\frac{V^{(2n)}_{\kappa}(0)}{H^{4-2n}},
\label{eq:effpotansatz:2}
\end{equation}
and with the initial conditions $c_{0,\kappa_{0}}=0$, $c_{2,\kappa_{0}}=-c_{4,\kappa_{0}}=-0.0001$, $\kappa_{0}/H=1$ 
and we take $c_{2n,\kappa_{0}}$ ($n\ge 3$) to be the coefficients of $\phi^{2n}$ 
in the SQED effective potential~\eqref{eq:sqedeffpot}. 
While the lower order terms ($n=0,1,2$) represent the classical Higgs-like potential,
the higher order local interactions  ($n\geq 3$) 
model accurately the scalar-photons interactions in the infrared, $k/a<H$. 
The running of the parameters of the effective potential~\eqref{eq:effpotansatz} are shown in 
figures~\ref{fig:run1s_extra}, \ref{fig:run1s}, \ref{fig:run2s}, \ref{fig:run3s} and~\ref{fig:run4s}.

\begin{figure}[H]
\begin{center}
\begin{tikzpicture}
\begin{groupplot}[group style={
            group size=1 by 4,
            vertical sep=0pt
                    },width=8.0cm,samples=100,xmin=-3.0,xmax=3.0,scaled ticks=false]
        
\nextgroupplot[
                        ymin=-0.00015,
                        ymax=0.00015,
                        yticklabels={-0.0001,0.0001},
                        ytick={-0.0001,0.0001},
                        axis x line=center,
                        axis y line=center, 
                        height=3.0cm,
                        xlabel={$\phi/H$},ylabel={$V_{\kappa}(\phi)/H^4$}
                       ]
 \addplot[color=red,smooth,line width=1.0pt,domain=-3.0:3.0]
 expression{0 - 0.00005*x^2 + 4.166666666666667e-6*x^4 + 4.6823290596437433e-7*x^6 - 3.4540304239548673e-9*x^8 + 3.601234603792603e-11*x^10 - 4.2852361230201943e-13*x^12 + 5.441709788974481e-15*x^14 - 7.170464679086218e-17*x^16 + 9.671765871009337e-19*x^18};\label{plots:plot1}
 \coordinate (top) at (rel axis cs:0,1);
 
 \nextgroupplot[
                        ymin=-0.06122,
                        ymax=-0.06092,
                        yticklabels={-0.0611,-0.0612},
                        ytick={-0.0611,-0.0612},
                        axis x line=none,
                        axis y line=center,
                        axis line style={-},
                        axis y discontinuity=crunch,
                        height=3.0cm,
                       ]
 \addplot[color=blue,smooth,line width=1.0pt,domain=-3.0:3.0]
 expression{-0.06118169553744896 - 0.000010788564967136963*x^2 + 9.911746429563457e-6*x^4 + 3.765470716342481e-7*x^6 - 2.7581100194819924e-9*x^8 + 3.701520038285352e-11*x^10 - 4.3659696733392793e-13*x^12 + 8.873733714974581e-15*x^14 - 1.8677726415029605e-17*x^16 + 1.5490400339630332e-18*x^18};\label{plots:plot2}
 
  \nextgroupplot[
                        ymin=-0.08725,
                        ymax=-0.08695,
                        yticklabels={-0.0871,-0.0872},
                        ytick={-0.0871,-0.0872},
                        axis x line=none,
                        axis y line=center,
                        axis line style={-},
                        axis y discontinuity=crunch,
                        height=3.0cm,
                       ]
 \addplot[color=brown,smooth,line width=1.0pt,domain=-1.0:1.0]
 expression{-0.0872306664000409 + 0.00023305249844211685*x^2 + 0.000017080440795417744*x^4 + 1.6002673056572365e-7*x^6 + 1.2911420263049911e-9*x^8 + 3.217420855931493e-11*x^10 + 6.167638219518589e-13*x^12 + 1.3585513852287637e-14*x^14 + 1.6413099893354603e-16*x^16 + 1.9248699389697434e-18*x^18};\label{plots:plot3}

 \nextgroupplot[
                        ymin=-0.1223,
                        ymax=-0.1220,
                        yticklabels={-0.1223,-0.1222},
                        ytick={-0.1223,-0.1222},
                        axis x line=none,
                        axis y line=center,
                        axis line style={-},
                        axis y discontinuity=crunch,
                        height=3.0cm,
                       ]
 \addplot[color=green,smooth,line width=1.0pt,domain=-1.0:1.0]
 expression{-0.12227584905877581 + 0.0014025395706853312*x^2 + 9.033821036085642e-6*x^4 + 1.159044741614859e-7*x^6 + 1.7608087783923315e-9*x^8 + 2.7974271541788688e-11*x^10 + 4.385000666879698e-13*x^12 + 6.474212244147695e-15*x^14 + 8.424859833487228e-17*x^16 + 8.085656227883485e-19*x^18};\label{plots:plot4}
\coordinate (bot) at (rel axis cs:1,0);
\end{groupplot}
\path (top|-current bounding box.north)--
      coordinate(legendpos)
      (bot|-current bounding box.north);

\matrix[
    matrix of nodes,
    anchor=north,
    draw,
    inner sep=0.2em,
    draw
  ]at([xshift=22ex,yshift=-30ex]legendpos)
  {
    \ref{plots:plot1}& $\kappa=\kappa_{0}=H$\\
    \ref{plots:plot2}& $\kappa=0.2H$\\
    \ref{plots:plot3}& $\kappa=0.1H$\\
    \ref{plots:plot4}& $\kappa=0$\\};
\end{tikzpicture}
\end{center}
\caption{Effective potential \eqref{eq:effpotansatz} for $\kappa=\kappa_{0}=H$, $\kappa=0.2H$, $\kappa=0.1H$ and $\kappa=0$. The potential becomes flat at $\kappa\simeq 0.184H$, the point of dynamical symmetry restoration.  For $\kappa < 0.184H$
the potential is positively curved  (concave) at the origin, implying that scalar particles become massive in the deep infrared.}
\label{fig:run8s}
\end{figure}
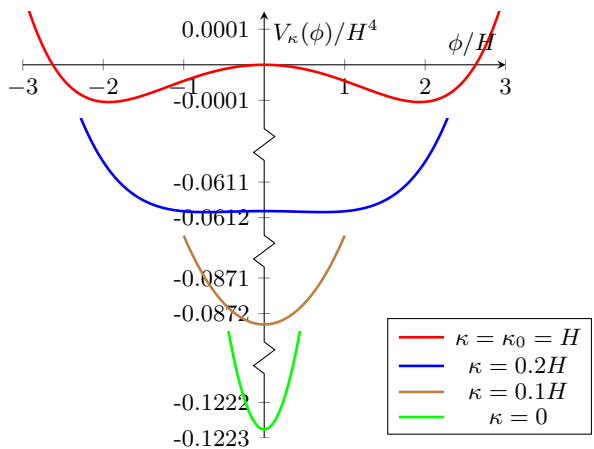

The fact that $c_{2,\kappa}$ becomes positive for small $\kappa$ means that the symmetry gets 
dynamically restored due to IR modes, this phenomena was first discussed by Ratra in~\cite{Ratra:1984yq}
and later in~\cite{Janssen:2009pb,Serreau:2011a,Lazzari:2013a,Serreau:2014a,Guilleux:2015a}. 
This can be seen in the flow of the minimum of the potential,
 $\bar{\phi}$, shown in figure~\ref{fig:run5s}. The running of the scalar field mass 
$\left.m_{\kappa}^{2}=V_{\kappa}''(\phi)\right|_{\phi=\bar{\phi}}$ (shown in figure~\ref{fig:run6s}) 
shows how the mass grows after the symmetry restoration point at which the mass is zero. 
It is interesting to see the running of the photon mass $m_{\gamma}^{2}=\Braket{e^{2}\phi^{2}}_{S}$ 
(see figure~\ref{fig:run7s}), calculated using the stochastic formalism late time expectation value~\cite{Starobinsky:1986a}. 
There we observe a non-vanishing photon mass, as shown in~\cite{Prokopec:2008b}, 
and a decrease of the mass in the IR. Note that the photon and scalar mass violate
the pertubative relation, $m_\gamma^2 m_\phi^2 = 3 e^2/(4\pi^2)$. This relation becomes particularly bad near the critical point
(at which the scalar mass vanishes). 
Finally we can observe the symmetry restoration of the effective 
potential in figure~\ref{fig:run8s} where we plot it for different values of the cutoff $\kappa$.

\begin{figure}[H]
\begin{center}
\begin{tikzpicture}
\begin{axis}[legend pos=south east,axis x line=center,axis y line=left,ylabel near ticks,width=8.5cm,height=5.8cm,={$\kappa/H$},xlabel={$\kappa/H$},ylabel={$\bar{\phi}/H$},xmin=0,xmax=1,ymin=0,ymax=2.0]
 \addplot[color=red,smooth,line width=1.0pt] table {min_0.0001.dat};
 \addplot[color=red,smooth,line width=1.0pt,domain=0.0:0.18393386151524657]  expression{0};
\end{axis}
\end{tikzpicture}
\end{center}
\caption{Running of the order parameter $\bar{\phi}$ in units of $H$. At $\kappa\simeq 0.184H$, $\bar\phi=0$ 
and the symmetry gets restored and stays restored for all $\kappa < 0.184H$.}
\label{fig:run5s}
\end{figure}
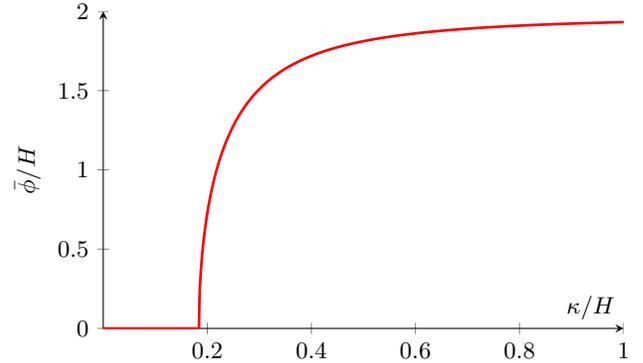

\begin{figure}[H]
\begin{center}
\begin{tikzpicture}
\begin{axis}[axis x line=center,axis y line=left,ylabel near ticks,width=8.5cm,height=5.8cm,xlabel={$\kappa/H$},ylabel={$m_{\kappa}^{2}/H^2$},xmin=0,xmax=1,ymin=0,ymax=0.003]
 \addplot[color=red,smooth,line width=1.0pt] table {mass_0.0001.dat};
\end{axis}
\end{tikzpicture}
\end{center}
\caption{Running of the scalar mass, $m_{\kappa}^{2}$ in units of $H^2$.}
\label{fig:run6s}
\end{figure}
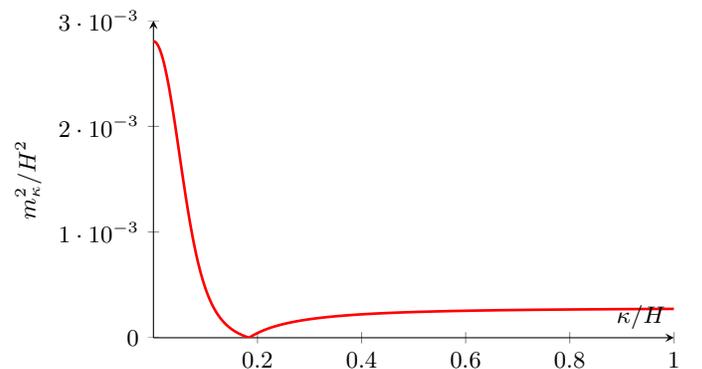

\begin{figure}[H]
\begin{center}
\begin{tikzpicture}
\begin{axis}[axis x line=center,axis y line=left,ylabel near ticks,width=8.5cm,height=5.8cm,xlabel={$\kappa/H$},ylabel={$m_{\gamma,\kappa}^{2}/H^2$},
xmin=0,xmax=1,ymin=0.7,ymax=1.35]
 \addplot[color=red,smooth,line width=1.0pt] table {massph_0.0001.dat};
\end{axis}
\end{tikzpicture}
\end{center}
\caption{Running of the photon mass, $m_{\gamma,\kappa}^{2}$ in units of $H^2$.}
\label{fig:run7s}
\end{figure}
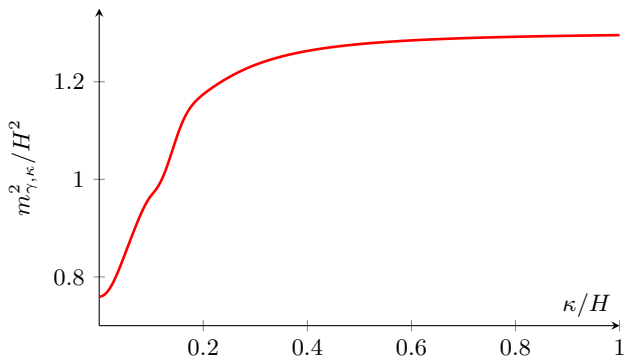

It is instructive to compare the results of this section with the results for real scalar field with a Higgs-like potential and 
without photons interactions. This model is analyzed in appendix~\ref{sec:a} 
(see also~\cite{Serreau:2014a,Guilleux:2015a}). 
There we see that the symmetry restoration point occurs at a lower energy scale than in the case when photons are included 
and therefore the mass for the scalar field is smaller. This means that photons enhance the symmetry restoration 
for the scalar field. We compare also our result for the mass with the results obtained using 
stochastic formalism~\cite{Lazzari:2013a} and find a very good agreement between two methods.

\subsection{\label{sec:3-2}Scalar QED stress-energy tensor}

\begin{figure}[hth]
\begin{center}
\begin{tikzpicture}
\begin{axis}[legend style={at={(1.04,0.98)},anchor=north west},axis x line=center,axis y line=center,width=7.0cm,height=4.9cm,xlabel={$\phi/H$},ylabel={$V_{\textrm{em},\kappa}\left(\phi\right)/H^4$},samples=100,xmin=-9.0,xmax=9.0,ymin=-0.08,ymax=0.08,every y tick scale label/.style={at={(0.5,1.03)},yshift=1pt,anchor=south west,inner sep=0pt}]
 \addplot[color=red,smooth,line width=1.0pt,domain=-9.0:9.0]
 expression{-0.0008960572993201664*x^2 - 0.000035771849192091835*x^4 + 7.023493589465614e-7*x^6 - 6.908060847909733e-9*x^8 + 9.003086509481507e-11*x^10 - 1.285570836906058e-12*x^12 + 1.9045984261410678e-14*x^14 - 2.868185871634487e-16*x^16 + 4.352294641954201e-18*x^18};
 \addplot[color=blue,smooth,line width=1.0pt,domain=-8.0:8.0]
 expression{-0.0008764511611463226*x^2 - 0.0000300266901918487*x^4 + 5.648190718637172e-7*x^6 - 5.516191631340317e-9*x^8 + 9.253760845230319e-11*x^10 - 1.3097670194753973e-12*x^12 + 3.1058253098822765e-14*x^14 - 7.470365601204698e-17*x^16 + 6.970619682563649e-18*x^18};
 \addplot[color=brown,smooth,line width=1.0pt,domain=-8.0:8.0]
 expression{-0.000754530200911403*x^2 - 0.000022858034043617135*x^4 + 2.400402698031084e-7*x^6 + 2.582291072375414e-9*x^8 + 8.043486502758018e-11*x^10 + 1.850269211732053e-12*x^12 + 4.754857292685391e-14*x^14 + 6.565151284745003e-16*x^16 + 8.661818350633035e-18*x^18};
 \addplot[color=green,smooth,line width=1.0pt,domain=-8.0:8.0]
 expression{-0.00016978690085820883*x^2 - 0.000030904686973576306*x^4 + 1.7385679482814582e-7*x^6 + 3.521617053631804e-9*x^8 + 6.993561576172831e-11*x^10 + 1.3154973175872386e-12*x^12 + 2.265964273977722e-14*x^14 + 3.3699515899865757e-16*x^16 + 3.638506363172641e-18*x^18};
 \legend{$\kappa=\kappa_{0}=H$,$\kappa=0.2H$,$\kappa=0.1H$,$\kappa=0$}
\end{axis}
\end{tikzpicture}
\end{center}
\caption{Stress-energy tensor effective potential \eqref{eq:effectivepotse} for $\kappa=\kappa_{0}=H$, $\kappa=0.2H$, $\kappa=0.1H$ and $\kappa=0$.}
\label{fig:run9s}
\end{figure}
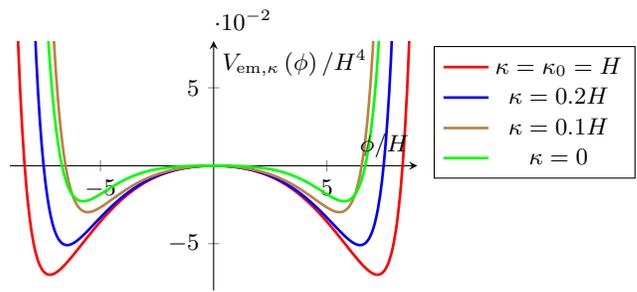

The stress-energy tensor $T_{\mu\nu}=-(2/\sqrt{-g})\delta S_{\rm SQED}/\delta g^{\mu\nu}$
 for the SQED action~\eqref{eq:sqedbare} is,
\begin{equation}
\begin{gathered}
T_{\mu\nu}=\left(\delta^{\alpha}_{\mu}\delta^{\beta}_{\nu}-\frac{1}{2}g_{\mu\nu}g^{\alpha\beta}\right)\left(\partial_{\alpha}\phi\partial_{\beta}\phi+e^2A_{\alpha}A_{\beta}\phi^{2}\right)\\
+\left(\delta^{\alpha}_{\mu}\delta^{\beta}_{\nu}g^{\gamma\delta}-\frac{1}{4}g_{\mu\nu}g^{\alpha\beta}g^{\gamma\delta}\right)F_{\alpha\gamma}F_{\beta\delta}\\
+\delta\xi\left(R_{\mu\nu}-\frac{1}{2}Rg_{\mu\nu}+g_{\mu\nu}\nabla_{\rho}\nabla^{\rho}-\nabla_{\mu}\nabla_{\nu}\right)\phi^{2}\\
-\frac{\delta\lambda}{4}\phi^{4}g_{\mu\nu}
\,.
\end{gathered}
\end{equation}
Approximating gravitational contributions with their background value,
$R_{\mu\nu}=dH^{2}g_{\mu\nu},R=d(d+1)H^{2}$ and
evaluating the photon and scalar contributions by making use of the stochastic framework developed 
in~\cite{Prokopec:2008b} one obtains, 
\begin{equation}
\begin{gathered}
\Braket{T_{\mu\nu}}=-g_{\mu\nu}\frac{3H^{4}}{8\pi^{2}}\left\{\left(-1+\gamma\right)z+\left(-2+\gamma\right)z^{2}\right.\\
\left.+\frac{1}{2}\left(z+z^{2}\right)\left[\Psi\left(\frac{3}{2}+\nu\right)+\Psi\left(\frac{3}{2}-\nu\right)\right]\right\}
,
\end{gathered}
\label{Tmn exact}
\end{equation}
where 
\begin{equation}
 \nu(z) = \sqrt{\frac14 - 2z}
\label{nu of z}
\end{equation}
To obtain the result~(\ref{Tmn exact}) we have performed dimensional regularization 
and made use of the same counterterms~\eqref{eq:cont1} and~\eqref{eq:cont2} as for the SQED 
effective action~\eqref{eq:sqedbare}. Upon expanding~(\ref{Tmn exact}) around $z=0$ 
we obtain, 
\begin{equation}
\begin{gathered}
\Braket{T_{\mu\nu}}=-g_{\mu\nu}\frac{3H^{4}}{8\pi^{2}}\left\{\left(-1+\gamma\right)z+\left(-2+\gamma\right)z^{2}\right.\\
\left.-\frac{1}{2}\left(z+z^{2}\right)\left[2\gamma-1+2\sum_{n=1}^{\infty}\zeta(2n+1)x^{2n}-\sum_{n=1}^{\infty}x^{n}\right]\right\},
\end{gathered}
\label{Tmn:3}
\end{equation}
where $x(z) = \frac12 - \sqrt{\frac14-2z}$.
Up to order $\phi^{18}$ Eq.~(\ref{Tmn:3}) reads,
\begin{equation}
\begin{gathered}
\Braket{T_{\mu\nu}}
=-g_{\mu\nu}\frac{3H^{4}}{8\pi^{2}}\left\{-\frac{1}{2}\left(\frac{e^{2}}{2H^{2}}\right)\phi^{2}\right.\\
-\frac{1}{2}\left(\frac{e^{2}}{2H^{2}}\right)^{2}\phi^{4}+\left[5-4\zeta(3)\right]\left(\frac{e^{2}}{2H^{2}}\right)^{3}\phi^{6}
 \\
+\left[24-20\zeta(3)\right]\left(\frac{e^{2}}{2H^{2}}\right)^{4}\phi^{8}\\
+\left[132-96\zeta(3)-16\zeta(5)\right]\left(\frac{e^{2}}{2H^{2}}\right)^{5}\phi^{10}\\
+\left[784-528\zeta(3)-144\zeta(5)\right]\left(\frac{e^{2}}{2H^{2}}\right)^{6}\phi^{12}\\
+\left[4896-3136\zeta(3)-1024\zeta(5)-64\zeta(7)\right]\left(\frac{e^{2}}{2H^{2}}\right)^{7}\phi^{14}\\
+\left[31680-19584\zeta(3)-7040\zeta(5)-832\zeta(7)\right]\left(\frac{e^{2}}{2H^{2}}\right)^{8}\phi^{16}\\
+\left[210496-126720\zeta(3)-48384\zeta(5)-7680\zeta(7)\right.\\
\left.\left.-256\zeta(9)\right]\left(\frac{e^{2}}{2H^{2}}\right)^{9}\phi^{18}+\mathcal{O}\left(\phi^{20}\right)\right\}.\label{eq:stressenergy}
\end{gathered}
\end{equation}
As required by de Sitter symmetry, this energy-momentum tensor is proportional to the metric tensor,
\begin{equation}
\Braket{T_{\mu\nu}}=-g_{\mu\nu}V_{\textrm{em}}\left(\phi\right),
\end{equation}
and the relation between $V_{\textrm{eff}}(z)$ and $V_{\textrm{em}}(z)$ is \cite{Prokopec:2008b}
\begin{equation}
V_{\textrm{eff}}(z)=\frac{3H^{4}}{8\pi^{2}}f(z),
\end{equation}
\begin{equation}
V_{\textrm{em}}(z)=\frac{3H^{4}}{8\pi^{2}}g(z),\label{eq:effectivepotse}
\end{equation}
with 
\begin{equation}
g(z)=\frac{1}{2}\left[zf'(z)-z-z^{2}\right]
. 
\label{g(z)}
\end{equation}
We take the same {\it Ansatz}~\eqref{eq:effpotansatz} for the effective potential $V_{\textrm{eff}}(\phi)$ but with the coupling constants of~\eqref{eq:stressenergy} as new initial conditions plus the initial conditions of the Higgs-like potential~\eqref{eq:sqedeffpot}.
This potential is plotted for different values of the cutoff $\kappa$ in figure~\ref{fig:run9s}. 
There we can observe a change in the minimum of the potential, although the value of the minimun lies in the region where the small field approximation starts to break down. We want to point out that the shape and minimum of the potential \eqref{eq:effectivepotse} can change significantly for a different choice of counterterms.

\section{\label{sec:4}Discussion}

We study the running under the nonperturbative renormalization group flow of scalar quantum electrodynamics (SQED) on
de Sitter space which at tree level exhibits spontaneous symmetry breaking due to a Higgs-like potential. 
This theory contains a charged scalar field canonically coupled a massless vector field.
We have performed our study by firstly integrating the photon, and then studied the renormalization group flow
in the resulting effective scalar field theory. 
Our results show that SQED on de Sitter exhibits dynamical symmetry restoration analogous to that of 
a pure scalar theory endowed with a symmetry breaking potential and studied in 
Refs.~\cite{Ratra:1984yq,Janssen:2009pb}, \cite{Lazzari:2013a,Serreau:2014a,Guilleux:2015a}. 
A detailed comparison with the work of Serreau and Guilleux~\cite{Serreau:2014a,Guilleux:2015a}
reveals that SQED exhibits a stronger symmetry restoration and a larger scalar field mass than the pure scalar theory.
Therefore, even though photons are conformal on de Sitter such that very few photons are generated,  
their quantum fluctuations facilitate symmetry restoration in the scalar sector. 
In Appendix~\ref{sec:a}
we compare the results obtained by RG methods with those obtained with the stochastic formalism 
in a pure scalar theory~\cite{Lazzari:2013a} and obtain an excellent agreement for the scalar mass. 

Even though the stochastic formalism and the renormalization group approach are at a first sight very different techniques, 
they yield identical results, provided stochastic theory is interpreted in the way advocated in Ref.~\cite{Lazzari:2013a}.
A closer look at the procedure advocated in~\cite{Lazzari:2013a} reveals that, approximating 
the effective action by the effective potential and studying the flow of the resulting truncated theory
mathematically corresponds to the identical procedure as advocated in~\cite{Lazzari:2013a} in the context of 
the stochastic theory of inflation.
Therefore, we can conclude that the effective potential obtained in~\cite{Lazzari:2013a}
is identical to effective potential $V_{\kappa\rightarrow 0}$ obtained by solving 
the renormalization group flow equation~(\ref{the flow of Vkappa}).
The advantage of the flow analysis is in that it allows for information on $V_{\kappa}$ at a finite value of 
$\kappa$ and in addition it allows for a systematic study of the running of higher derivative operators,
the lowest order one being $\sim Z_\kappa(\phi)\partial_\mu \phi \partial_\mu \phi g^{\mu\nu}$. 
The advantage of the stochastic approach is that analytical results can be obtained for 
$V_{\rm eff}$, both in the small field regime (mass term) and in the large field regime (asymptotic expansion),
for details see~\cite{Lazzari:2013a}.

 Up to now we have not discussed whether one could assign a physical meaning to the effective potential 
$V_\kappa(\phi)$ at a finite value of $\kappa$. Note that the nonperturbative effective potential exhibits similar features 
to the perturbative effective (Hubble) potential studied in Ref.~\cite{Janssen:2009pb}. 
Namely, the Hubble potential exhibits de Sitter breaking induced both by the scale dependent counterterms and 
by the infrared effects (as time passes more and more infrared modes get populated, which is manifested as a
de Sitter breaking time
dependence in the effective potential). 
Analogously, as $\kappa$ gets smaller and smaller, more and more infrared modes get integrated out.
Therefore, if one interprets $\kappa$ as the physical momentum cutoff scale below which the modes are mostly unpopulated,
then one can make the following replacement, $\kappa\rightarrow k_0/a$, with $k_0$ being the infrared 
comoving cutoff scale (defined as the physical momentum at the time when $a_0=a(\eta_0)=1$) which is naturally expected 
to be of the order or smaller than the Hubble scale, $k_0\lesssim H$. With this one gets,
\begin{equation}
 V_\kappa(\phi)\rightarrow V_{k_0/a(\eta)}(\phi)
\,.
\label{interpretation V_kappa}
\end{equation}
With this in mind one can reinterpret figure~\ref{fig:run8s} as a sequence of snapshots at times 
$t(\kappa) = (1/H)\ln(k_0/\kappa)\simeq (1/H)\ln(H/\kappa)$ of the effective potential. 
This suggests that the symmetry restoration occurs quite quickly,
within a few e-foldings of inflation (for the chosen set of parameters). This is in accordance with the perturbative study 
of Ref.~\cite{Janssen:2009pb}. The principal advantage of the current work is that the effective potential 
presented here is truly nonperturbative, and holds true both at early as well as at late times.
Indeed, while the Hubble effective potential of Ref.~\cite{Janssen:2009pb} never settles to a (de Sitter invariant) state, 
the nonperturbative effective potential $V_\kappa$ reaches quite quickly a de Sitter invariant state
(which is formally reached in the limit when $\kappa\rightarrow 0$). 

This way of interpreting $V_\kappa$ 
can have important consequences of our understanding of cosmological perturbations in theories when quantum corrections
to the tree potential are large, especially when one keeps in 
mind that the small field expansion
is the relevant regime~\cite{Bilandzic:2007nb} of the effective potential during inflation.
Namely, it is usually assumed that the only dependence on time in the potential is indirect
{\it via} the time dependence of the inflaton field, $\phi=\phi(t)$.
This then gives for {\it e.g.}
the principal slow roll parameter, $\epsilon = -\dot H/H^2 \approx (M_{\rm P}^2/2)(V^\prime(\phi)/V)^2$, where the notation $\dot H$ stands now for time derivative i.e. $\mathrm{d}H/\mathrm{d}t$.
But, replacing in this formula $V$ by $V_{\kappa=H/a}$ can lead to a significant change in $\epsilon$ and  
in a particularly large change in $\epsilon$'s  dependence on time, which is reflected in the second 
slow roll parameter $\eta=-\dot\epsilon/(\epsilon H)$. 
A detailed investigation of this important question is postponed to future work.

As a final remark we note that one could study scalar field theories with extended symmetries, important examples of which are 
the standard model Higgs sector (in which the symmetry group is $SU(2)$) 
and various grand unified model Higgs sectors. 
One can accomplish this by using the same renormalization group techniques 
(see~\cite{Serreau:2014a,Guilleux:2015a}).

\bigskip

\appendix
\section{\label{sec:a}Scalar field theory}

In this appendix we calculate the flow of a spontaneously broken scalar field theory without photons interactions, 
which is known to exhibit dynamical symmetry restoration driven by the enhanced infrared field 
fluctuations~\cite{Ratra:1984yq,Janssen:2009pb}, \cite{Lazzari:2013a,Serreau:2014a,Guilleux:2015a}.
In order to study symmetry restoration we take the following {\it Ansatz} for the effective potential,
\begin{equation}
\begin{gathered}
V_{\kappa}(\phi)=H^4\sum_{n=0}^{\infty}\frac{c_{2n,\kappa}}{(2n)! }\left(\frac{\phi^2}{H^2}\right)^{n}
,
\label{eq:higgs}
\end{gathered}
\end{equation}
where the couplings are defined as,
\begin{equation}
c_{2n,\kappa}=\frac{V^{(2n)}_{\kappa}(0)}{H^{4-2n}}
.
\label{eq:higgs:2}
\end{equation}

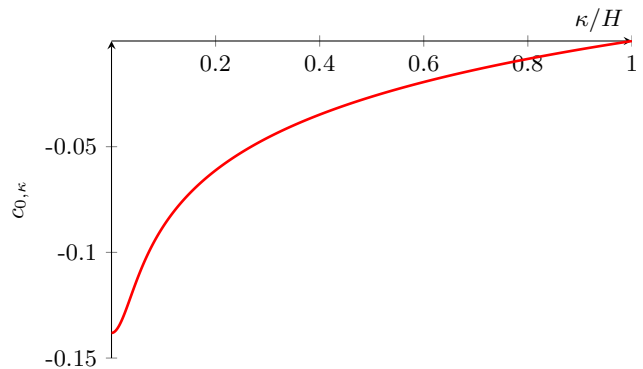
\begin{figure}[H]
\begin{center}
\begin{tikzpicture}
\begin{axis}[axis x line=center,axis y line=left,ylabel near ticks,ytick={-0.05,-0.1,-0.15},yticklabels={-0.05,-0.1,-0.15},width=8.5cm,height=5.8cm,xlabel={$\kappa/H$},ylabel={$c_{0,\kappa}$},xmin=0,xmax=1,ymin=-0.15,ymax=0.0]
 \addplot[color=red,smooth,line width=1.0pt] table {con0,18_0.0001.dat};
\end{axis}
\end{tikzpicture}
\end{center}
\caption{Running of the constant $c_{0,\kappa}$ 
defined in~(\ref{eq:higgs}--\ref{eq:higgs:2}).}
\label{fig:run10s_extra}
\end{figure}

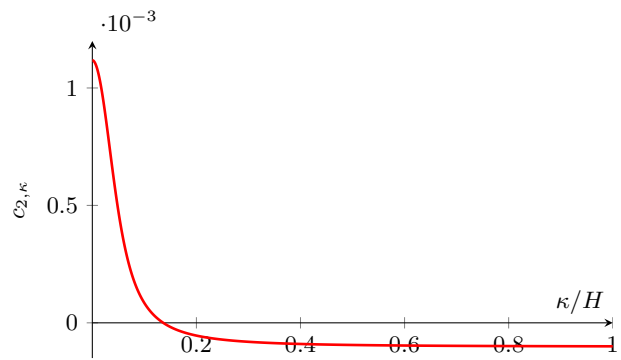
\begin{figure}[H]
\begin{center}
\begin{tikzpicture}
\begin{axis}[axis x line=center,axis y line=left,ylabel near ticks,width=8.5cm,height=5.8cm,xlabel={$\kappa/H$},ylabel={$c_{2,\kappa}$},xmin=0,xmax=1,ymin=-0.00015,ymax=0.0012]
 \addplot[color=red,smooth,line width=1.0pt] table {con2,18_0.0001.dat};
\end{axis}
\end{tikzpicture}
\end{center}
\caption{Running of the quadratic coupling constant $c_{2,\kappa}$ (mass term) 
defined in~(\ref{eq:higgs}--\ref{eq:higgs:2}). When compared with figure~\ref{fig:run1s}, 
one sees that the symmetry restoration occurs earlier (for a larger value of $\kappa$) in SQED.}
\label{fig:run10s}
\end{figure}
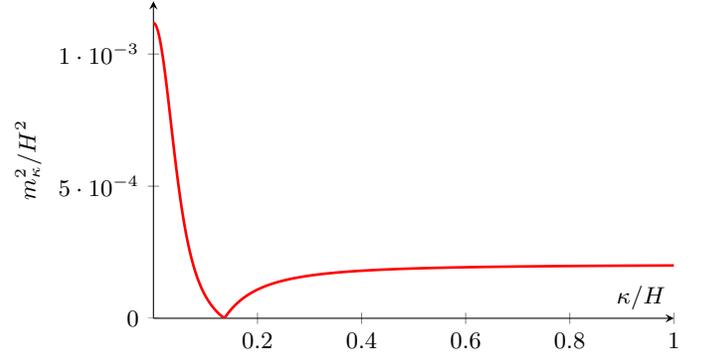
\begin{figure}[H]
\begin{center}
\begin{tikzpicture}
\begin{axis}[axis x line=center,axis y line=left,ylabel near ticks,width=8.5cm,height=5.8cm,xlabel={$\kappa/H$},ylabel={$m_{\kappa}^{2}/H^2$},xmin=0,xmax=1,ymin=0.0,ymax=0.0012]
 \addplot[color=red,smooth,line width=1.0pt] table {mass,18_0.0001.dat};
\end{axis}
\end{tikzpicture}
\end{center}
\captionof{figure}{Flow of the scalar mass $m_{\kappa}^{2}$ defined from~(\ref{eq:higgs}--\ref{eq:higgs:2}).
}
\label{fig:run12s}
\end{figure}
\begin{figure}[H]
\begin{center}
\begin{tikzpicture}
\begin{groupplot}[group style={
            group size=1 by 4,
            vertical sep=0pt
                    },width=8.0cm,samples=100,xmin=-4.0,xmax=4.0,scaled ticks=false]
        
\nextgroupplot[
                        ymin=-0.0002,
                        ymax=0.0001,
                        yticklabels={-0.0001,0.0001},
                        ytick={-0.0001,0.0001},
                        axis x line=center,
                        axis y line=center, 
                        height=3.0cm,
                        xlabel={$\phi/H$},ylabel={$V_{\kappa}(\phi)/H^4$},
                        legend style={at={(1.04,0.98)},anchor=north west}
                       ]
 \addplot[color=red,smooth,line width=1.0pt,domain=-4.0:4.0]
 expression{0 - 0.00005*x^2 + 4.166666666666667e-6*x^4};\label{plots:plot1}
 \coordinate (top) at (rel axis cs:0,1);
 
 \nextgroupplot[
                        ymin=-0.06127,
                        ymax=-0.06097,
                        yticklabels={-0.0612,-0.0611},
                        ytick={-0.0612,-0.0611},
                        axis x line=none,
                        axis y line=center,
                        axis line style={-},
                        axis y discontinuity=crunch,
                        height=3.0cm,
                       ]
 \addplot[color=blue,smooth,line width=1.0pt,domain=-4.0:4.0]
 expression{-0.06118704593013817 - 0.000027186377857839585*x^2 + 4.151888159554945e-6*x^4 + 1.2216226237085552e-11*x^6 - 1.1247551044716519e-14*x^8 + 1.0925247441032644e-17*x^10 - 1.0904224851707947e-20*x^12 + 1.1006667894593047e-23*x^14 - 1.1104408304196194e-26*x^16 + 1.1292460973343655e-29*x^18};\label{plots:plot2}
 
  \nextgroupplot[
                        ymin=-0.08755,
                        ymax=-0.08725,
                        yticklabels={-0.0874,-0.0875},
                        ytick={-0.0874,-0.0875},
                        axis x line=none,
                        axis y line=center,
                        axis line style={-},
                        axis y discontinuity=crunch,
                        height=3.0cm,
                       ]
 \addplot[color=brown,smooth,line width=1.0pt,domain=-2.0:2.0]
 expression{-0.08750147086668095 + 0.00004214291070707838*x^2 + 3.952521091589782e-6*x^4 + 5.979762079463547e-10*x^6 - 1.6456228044548834e-12*x^8 + 3.9617613107471874e-15*x^10 - 6.818168070477252e-18*x^12 + 7.261012741636541e-22*x^14 + 4.7178525920067e-23*x^16 - 2.3598835160533286e-25*x^18};\label{plots:plot3}

 \nextgroupplot[
                        ymin=-0.13817,
                        ymax=-0.13787,
                        yticklabels={-0.1380,-0.1381},
                        ytick={-0.1380,-0.1381},
                        axis x line=none,
                        axis y line=center,
                        axis line style={-},
                        axis y discontinuity=crunch,
                        height=3.0cm,
                       ]
 \addplot[color=green,smooth,line width=1.0pt,domain=-2.0:2.0]
 expression{-0.13811245530651378 + 0.0005589220223026108*x^2 + 1.1629056658797564e-6*x^4 + 4.252144087376925e-9*x^6 + 1.5302757098558163e-11*x^8 + 3.550718016095647e-14*x^10 - 1.0951375807109471e-16*x^12 - 2.2273444565206192e-18*x^14 - 1.6699494847319752e-20*x^16 - 1.0040462457566735e-22*x^18};\label{plots:plot4}
\coordinate (bot) at (rel axis cs:1,0);
\end{groupplot}
\path (top|-current bounding box.north)--
      coordinate(legendpos)
      (bot|-current bounding box.north);

\matrix[
    matrix of nodes,
    anchor=north,
    draw,
    inner sep=0.2em,
    draw
  ]at([xshift=22ex,yshift=-30ex]legendpos)
  {
    \ref{plots:plot1}& $\kappa=\kappa_{0}=H$\\
    \ref{plots:plot2}& $\kappa=0.2H$\\
    \ref{plots:plot3}& $\kappa=0.1H$\\
    \ref{plots:plot4}& $\kappa=0$\\};
\end{tikzpicture}
\end{center}
\caption{Higgs effective potential~\eqref{eq:higgs} for $\kappa=\kappa_{0}=H$, 
$\kappa=0.2H$, $\kappa=0.1H$ and $\kappa=0$.}
\label{fig:run13s}
\end{figure}
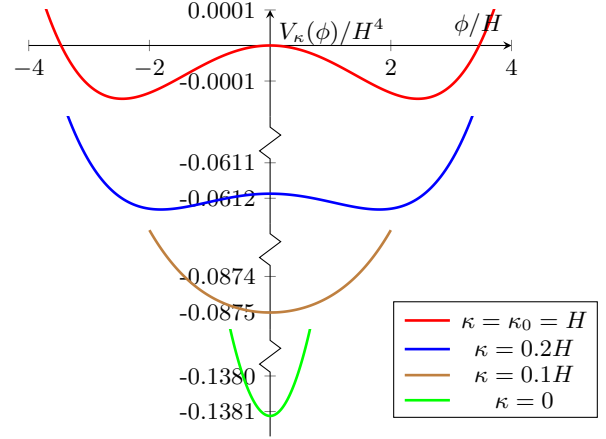

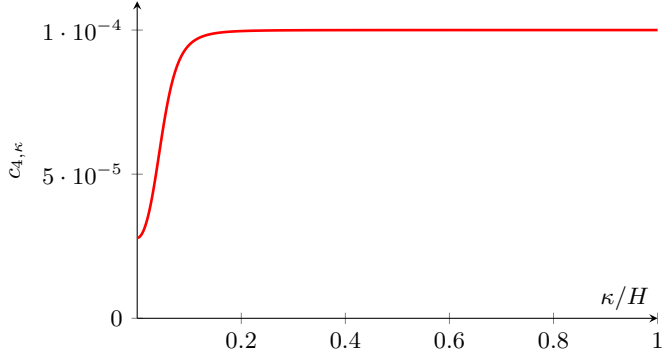
\begin{figure}[H]
\begin{center}
\begin{tikzpicture}
\begin{axis}[axis x line=center,axis y line=left,ylabel near ticks,width=8.5cm,height=5.8cm,xlabel={$\kappa/H$},ylabel={$c_{4,\kappa}$},xmin=0,xmax=1,ymin=0,ymax=0.00011]
 \addplot[color=red,smooth,line width=1.0pt] table {con4,18_0.0001.dat};
\end{axis}
\end{tikzpicture}
\end{center}
\caption{Running of the quartic coupling constant $c_{4,\kappa}$ defined in~(\ref{eq:higgs}--\ref{eq:higgs:2}).}
\label{fig:run11s}
\end{figure}
Taking as initial conditions, $c_{0,\kappa_{0}}=0$, $c_{2,\kappa_{0}}=-c_{4,\kappa_{0}}=-0.0001$, $\kappa_{0}=H$  
and $c_{2n,\kappa_{0}}=0$ ($n=3,4,\dots$), the effective potential is plotted for several values of $\kappa$ in figure \ref{fig:run13s}, 
where we included terms up to order $\phi^{18}$. 
We plot as well the running of $c_{0,\kappa}$ (figure \ref{fig:run10s_extra}), $c_{2,\kappa}$ 
(figure \ref{fig:run10s}), $c_{4,\kappa}$ (figure \ref{fig:run11s}) and the mass $m_{\kappa}^{2}$ (figure \ref{fig:run12s}). 
In this case we see that the symmetry gets restored at a lower energy scale $\kappa$ compared with the same potential 
with photons interactions included, see figures~\ref{fig:run5s} and~\ref{fig:run6s}.
This is the same result as Guilleux and Serreau obtained in~\cite{Serreau:2014a,Guilleux:2015a}, although they take different initial conditions,
thus obtaining different values for the mass and the symmetry restoration point.

In order to compare the RG approach with the stochastic formalism we can look at {\it e.g.} the scalar mass. The mass obtained in 
the limit $\kappa=0$ is $m^{2}=\num{1.11805e-3} H^{2}$ is to be compared with the late-time mass using the stochastic 
formalism~\cite{Lazzari:2013a}, $m^{2}=\num{1.11868e-3} H^{2}$. Therefore, there is an excellent
 agreement between the masses using both methods.

\bibliography{main}

\end{document}